\documentclass[sigconf]{acmart}

\copyrightyear{2027}
\acmYear{2027}
\setcopyright{cc}
\setcctype{by}
\acmConference[ASPDAC '27]{32nd Asia and South Pacific Design Automation Conference}{January 25--28, 2027}{Tokyo, Japan}
\acmBooktitle{32nd Asia and South Pacific Design Automation Conference (ASPDAC '27), January 25--28, 2027, Tokyo, Japan}
\acmDOI{10.1145/3840404.3848173}
\acmISBN{979-8-4007-2923-2/2027/01}

\usepackage[utf8]{inputenc}
\usepackage{textcomp}
\usepackage{pifont}
\usepackage{makecell}

\usepackage{url}
\newcommand{\cnum}[2][black]{\ifcase#2 \or\ding{182}\or\ding{183}\or\ding{184}\or\ding{185}\fi}

\newcommand{\cmark}{\ding{51}}
\newcommand{\xmark}{\ding{55}}

\AtBeginDocument{\providecommand\BibTeX{{\normalfont B\kern-0.5em{\scshape i\kern-0.25em b}\kern-0.8em\TeX}}}

\newcommand\blfootnote[1]{\begingroup
  \renewcommand\thefootnote{}\addtocounter{footnote}{-1}\endgroup
}

\usepackage{enumitem}
\setlist[itemize]{leftmargin=0.5cm}

\begin{document}

\title{MX-like Tensor Core for LLM Inference with NPU Integration via
Hardware-Software Co-Design Exploration}

\author{Zijun Zhao}
\authornote{Both authors contributed equally to this work.}
\orcid{0009-0008-5197-8507}
\email{zijun.zhao@student.kuleuven.be}
\affiliation{\institution{KU Leuven}
  \city{Leuven}
  \country{Belgium}}

\author{Ruichu Ao}
\authornotemark[1]
\orcid{0009-0004-8265-1495}
\email{ruichu.ao@student.kuleuven.be}
\affiliation{\institution{KU Leuven}
  \city{Leuven}
  \country{Belgium}}

\author{Stef Cuyckens}
\orcid{0009-0002-4526-7780}
\email{stef.cuyckens@kuleuven.be}
\affiliation{\institution{KU Leuven}
  \city{Leuven}
  \country{Belgium}}

\author{Marian Verhelst}
\orcid{0000-0003-3495-9263}
\email{marian.verhelst@kuleuven.be}
\affiliation{\institution{KU Leuven}
  \city{Leuven}
  \country{Belgium}}

\author{Chao Fang}
\authornote{Corresponding author.}
\orcid{0000-0003-3430-1189}
\email{chao.fang@kuleuven.be}
\affiliation{\institution{KU Leuven}
  \city{Leuven}
  \country{Belgium}}

\renewcommand{\shortauthors}{Zhao and Ao, et al.}

\begin{abstract}
Block-scaled quantization formats such as OCP Microscaling (MX) and NVFP4 have been widely adopted
as efficient data formats for LLM inference.
However, the design space of these formats spanned by shared scale, weight, and activation
format combinations has not been systematically explored,
and the joint accuracy--silicon tradeoff across this
space remains uncharacterized, leaving the true Pareto-optimal configurations for NPU-based edge LLM deployment an open question.
To bridge this gap,
we conduct a joint hardware–software (HW–SW) co-design exploration of the above underexplored MX-like design space with three innovations.
1) We define and systematically characterize the full 8-bit scale format family within the MX-like design space, jointly with asymmetric weight-activation precision, identifying configurations that outperform the MX and NVFP4-like (without per-tensor
scaling, PTS) baselines.
2) We build a per-format specialized MAC datapath template that enables fair hardware cost evaluation across the space.
3) We develop an automated co-design flow, integrated into the open-source SNAX NPU platform, that enables joint HW-SW analysis across the design space.
Overall, our co-design exploration reveals that scale format matters
only under aggressive FP4 quantization, yielding a case-oriented design
guideline for both scale format and asymmetric weight-activation
precision.
We further identify UE5M3 as a favorable accuracy–hardware tradeoff for the FP4 regime, reducing perplexity degradation by up to
25.3\% relative to the NVFP4-like baseline,
at a cost of only $+0.06$\% area and $+0.49$\% power overhead.
\end{abstract}



\maketitle

\begin{figure}[!t]
\centering
\includegraphics[width=\linewidth]{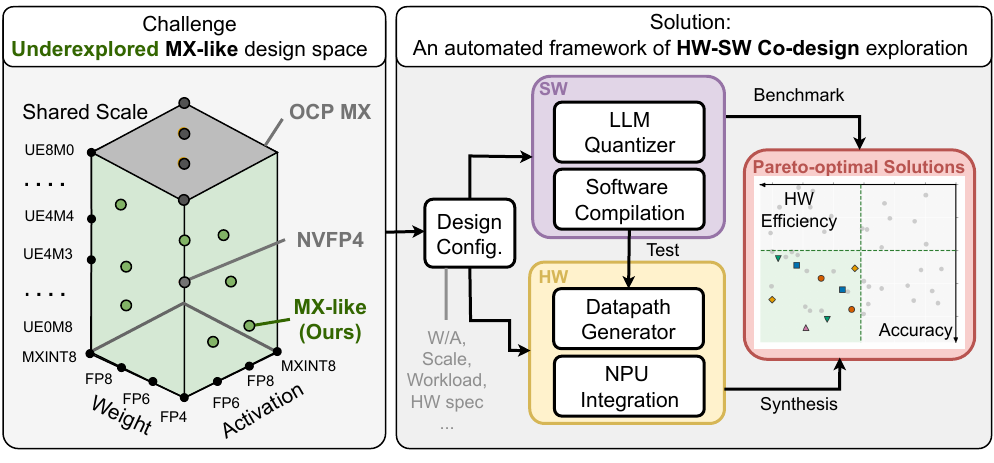}
\Description{Left: a cube spanned by the shared scale format (UE8M0 to UE0M8),
the weight format, and the activation format (MXINT8, FP8, FP6, and FP4). OCP MX
and NVFP4 are single points on the cube, and the MX-like design space of this
work fills the region between them. Right: the automated co-design framework.
A design configuration drives a software side (LLM quantizer and software
compilation) and a hardware side (datapath generator and NPU integration).
The benchmark and synthesis results give a plot of accuracy against hardware
efficiency that shows the Pareto-optimal solutions.}
\vspace{-1em}
\caption{Overview of our proposed MX-like design space (left) and our automated HW-SW
co-design framework that explores it to find Pareto-optimal configurations (right).}
\vspace{-1em}
\label{fig:intro}
\end{figure}

\section{Introduction}
Large language models (LLMs)
have demonstrated remarkable performance across a wide range of tasks, from natural language understanding to code generation~\cite{grattafiori2024llama3herdmodels}. However, this growing capability comes with a corresponding growth in model size~\cite{Bahri2024scaling}, demanding ever more memory and computational resources. This makes efficient LLM deployment a significant challenge, an issue that becomes even more pressing when deployment targets shift from datacenter GPUs~\cite{ma2026challenges} toward edge NPUs~\cite{edgellm2025survey}.

Block-scaled quantization~\cite{dai2021vs,rouhani2023microscaling,nvidia2026pretraininglargelanguagemodels,anda}
addresses this by quantizing small groups within the tensor and assigning a shared scale to each group to accommodate its local distribution. This allows high-precision values to be stored and computed in low-bitwidth datatypes, reducing memory footprint and power consumption while preserving accuracy.
As shown in Fig.~\ref{fig:intro}, the OCP MX~\cite{ocp_mx_spec_2023} specification and NVIDIA's
NVFP4~\cite{nvidia2026pretraininglargelanguagemodels} mark two
industry endpoints in the resulting accuracy-hardware tradeoff. MX adopts an unsigned power-of-two block
scale (UE8M0) for its prime hardware efficiency, while NVFP4 instead adopts a fractional UE4M3 block
scale to improve quantization granularity.

However, MX and NVFP4 each commit to a single point in the scale format space, UE8M0 and UE4M3, respectively. On the algorithmic side, prior works~\cite{rouhani2023microscaling, fasoli2026finerbetterlimitsmicroscaling} have only
explored a small part of
the MX-like design space in Fig.~\ref{fig:intro}: \cite{rouhani2023microscaling} sweeps element formats across INT8, FP8, FP6, and FP4 but fixes the scale format to E8M0, while \cite{fasoli2026finerbetterlimitsmicroscaling} explores fractional scale formats but restricts the analysis to FP4, leaving the joint interaction between scale format and element format not yet fully characterized. This uncertainty is compounded by a further consideration: activation and weight tensors exhibit different distributions~\cite{xiao2023smoothquant,yao2023zeroquantv2exploringposttrainingquantization}. Activations
are heavy-tailed with significant outliers,
whereas weights follow a near-Gaussian profile, suggesting that asymmetric weight-activation configurations could be beneficial.
Such asymmetry has mainly been studied in weight-only quantization at per-tensor and per-token granularity~\cite{xiao2023smoothquant,yao2023zeroquantv2exploringposttrainingquantization}, but with limited attention in the block-scaled setting.

On the hardware side, the silicon cost surface over the MX-like
format space remains unquantified. Prior MX-compatible MAC datapaths fix the scale to
UE8M0 with symmetric (activation, weight) operands, either as
a single FP8 implementation~\cite{mxdotp,lutz2024arithmetic} or as a runtime
precision-scalable unit that reuses the hardware for all MX element formats~\cite{cuyckens2025efficient,cuyckens2024precision}, making
the intrinsic cost of one format combination unobservable. Besides, fractional scales introduce a scale-mantissa multiplier and the operand pairs change the datapath width throughout the whole datapath, which couples the cost of one format axis to the other two.
Measuring this space fairly requires a per-format combination datapath under one uniform micro-architecture.  \begin{table}[!t]
\centering
\caption{Comparison with prior works}
\label{tab:dse_comparison}
\renewcommand{\arraystretch}{1.1}
\setlength{\tabcolsep}{2.5pt}
\resizebox{\columnwidth}{!}{\begin{tabular}{l | c c c c c}
\toprule
\makecell[l]{Related Works} &
DSE &
\makecell{Scale} &
\makecell{Asym.$^{\dagger}$\\(A,\,W)} &
\makecell{Silicon\\cost} &
\makecell{NPU\\integ.} \\
\midrule
Rouhani \textit{et al.}~\cite{rouhani2023sharedmicroexponentslittleshifting}        & HW+SW   & ExM0        & $\sim$       & \cmark & \xmark \\
FP4-All-the-Way~\cite{chmiel2025fp4wayfullyquantized}          & SW$^{FP4}$      & E8M0--E1M6 & \xmark  & \xmark & \xmark \\
Fasoli \textit{et al.}~\cite{fasoli2026finerbetterlimitsmicroscaling}      & SW$^{FP4}$      & UE5M3 & \xmark       & $\sim$ & \xmark \\
FGMP~\cite{hsia2025fgmp}                          & \xmark  & UE4M3       & $\sim$ & \cmark & \xmark \\
Cuyckens \textit{et al.}~\cite{cuyckens2024precision} & \xmark  & UE8M0       & \xmark       & \cmark & \cmark \\
MXDotP~\cite{mxdotp}                      & \xmark  & UE8M0       & \xmark       & \cmark & \cmark \\
\midrule
\textbf{Our work$\ddagger$}
& \textbf{HW+SW}
& \textbf{UE8M0--UE4M4}
& \textbf{\cmark}
& \textbf{\cmark}
& \textbf{\cmark} \\
\bottomrule
\end{tabular}}
\raggedright
\footnotesize
$\sim$: partial coverage.
$^{\dagger}$: \xmark{} in Asym.\ denotes symmetric-only.
$^{FP4}$: FP4 regime only.\\
$\ddagger$: Code is available at \url{https://github.com/KULeuven-MICAS/MX-Like-for-LLM-Inference}.

\end{table}

Moreover, the above two gaps are not independent. Examining accuracy and hardware cost in isolation is insufficient,
since a format that excels on one may be suboptimal on the other. Algorithmic studies~\cite{fasoli2026finerbetterlimitsmicroscaling,rouhani2023microscaling} treat hardware cost with varying and often limited rigor rather than systematic silicon evaluation, while prior hardware designs~\cite{cuyckens2024precision,mxdotp} characterize silicon cost for a fixed format. As a result, the joint accuracy-silicon Pareto frontier for edge NPU deployment remains unmapped.

In this work, as shown in Fig.~\ref{fig:intro}(left), we address these gaps with a joint hardware–software (HW–SW) co-design exploration of the MX-like design space, spanned by the scale, weight, and activation formats, that lies between MX and NVFP4. To fairly expose the cost of each format combination despite the coupling across format axes, we build a per-format specialized datapath under one uniform micro-architecture, and integrate it into an automated flow for joint accuracy-silicon analysis targeting edge NPU deployment. Our contributions are as follows:

\begin{itemize}
   \item \textbf{MX-like design space characterization:} We identify a
   previously overlooked region within the MX-like design space,
   spanning intermediate scale formats and asymmetric
   weight-activation precision, and systematically characterize it
   to reveal configurations that outperform both the MX and the NVFP4-like baselines.

    \item \textbf{Parameterized MAC datapath template:}
    We design a parameterized template for an MX-compatible datapath to
    support MX-like format combinations and evaluate the variation of hardware cost across the design space.

    \item \textbf{Automated co-design flow:} We provide an automated end-to-end flow encompassing hardware generation, system integration, and numerical verification, enabling efficient design space exploration (DSE) and HW-SW co-analysis that identifies practical configurations under application-specific accuracy and hardware budget constraints.
\end{itemize}

After constructing the Pareto front for hardware efficiency and software accuracy, we find that the UE8M0 scale is preferred above A6W6 while the UE5M3 scale dominates FP4 quantization, with asymmetric configurations and narrow-exponent element formats offering further gains. Together, these findings provide case-oriented guidelines for selecting block-scaled quantization configurations under application-specific accuracy and hardware constraints.

\section{BACKGROUND AND MOTIVATION}

Table~\ref{tab:dse_comparison} positions our work against the prior works along the
five axes that reveal the gap.
To the best of our knowledge, no prior art spans all five.
We elaborate on the following aspects.

\textbf{Quantization algorithms for LLMs: }Recent algorithmic studies refine the block-scaled recipe in two
directions. The first direction tunes the scale format itself.
~\cite{chmiel2025fp4wayfullyquantized} sweeps over FP8-typed scales in FP4 training and identifies fractional
E4M3 scales as the accuracy
optimum.
Fasoli et al.~\cite{fasoli2026finerbetterlimitsmicroscaling}
propose UE5M3 scale for FP4 inference.
The second direction diversifies the representation.
Per-block schemes such as AMXFP4~\cite{lee2024amxfp4} and
BlockDialect~\cite{jang2025blockdialect} establish the value of asymmetric,
block-specific formats.
The strongest hardware evidence in this line consists of spot
checks of one proposal against one baseline: a single-point area
comparison of UE5M3 scale
logic~\cite{fasoli2026finerbetterlimitsmicroscaling}, a MAC
synthesis placing AMXFP4 about 10\% over
MXFP4~\cite{lee2024amxfp4}, and a MAC-level match against
FP4~\cite{jang2025blockdialect}.
None of these studies
map how cost varies across the scale family, how it
couples with the element formats, or whether the FP4-regime
findings hold at higher precision, motivating our extension of the design space.

\textbf{MX-compatible MAC and tensor core: } Hardware support for block-scaled formats follows three strategies.
1) Fixed-function units implement one format well. Examples include a
fused FP8 dot product unit~\cite{lutz2024arithmetic} and an ISA
extension on the Snitch cluster~\cite{mxdotp}.
2) Runtime hardware-reuse makes the datapath precision-scalable: sub-word
parallelism reuses the 2-bit multipliers and adders across
all six MX element
formats~\cite{cuyckens2024precision,cuyckens2025efficient}.
3) Runtime replication instead carries parallel units for each
supported pair. FGMP~\cite{hsia2025fgmp} places four dot-product
units per lane to mix FP4 and FP8 blocks, at 3.5$\times$ the
datapath area of a single-format unit.
These designs~\cite{lutz2024arithmetic,mxdotp,cuyckens2024precision,cuyckens2025efficient,hsia2025fgmp} differ in how they trade efficiency for format
coverage, but they leave the same two dimensions untouched. Each
fixes the block scale to a single format, UE8M0 or UE4M3, and none supports fully asymmetric operation analysis. As a result, prior designs cannot realize or evaluate the broader scale-format and asymmetric space explored in this work, which motivates a datapath template specialized per configuration to expose the hardware cost of each one.

\textbf{Joint accuracy-silicon exploration: }Joint exploration of accuracy and silicon cost has been applied to
several axes of the quantization design space.
A few works co-explore accuracy and
silicon cost, each along one axis: quantization bit-widths in
VS-Quant~\cite{dai2021vs}, block-scaling structure in shared
microexponents~\cite{rouhani2023sharedmicroexponentslittleshifting} which seeded
the MX standard~\cite{ocp_mx_spec_2023}, and number systems with MAC parameters
in a recent sweep~\cite{crafton2025pareto}. None of these studies varies the exponent-mantissa split of the block scale format
itself, nor pairs it with a full range of asymmetric weight-activation operands.

\section{Design space exploration}

\subsection{MX-like Design Space Characterization}
\label{subsec:design_space}

\begin{figure}[!t]
\centering
\includegraphics[width=\linewidth]{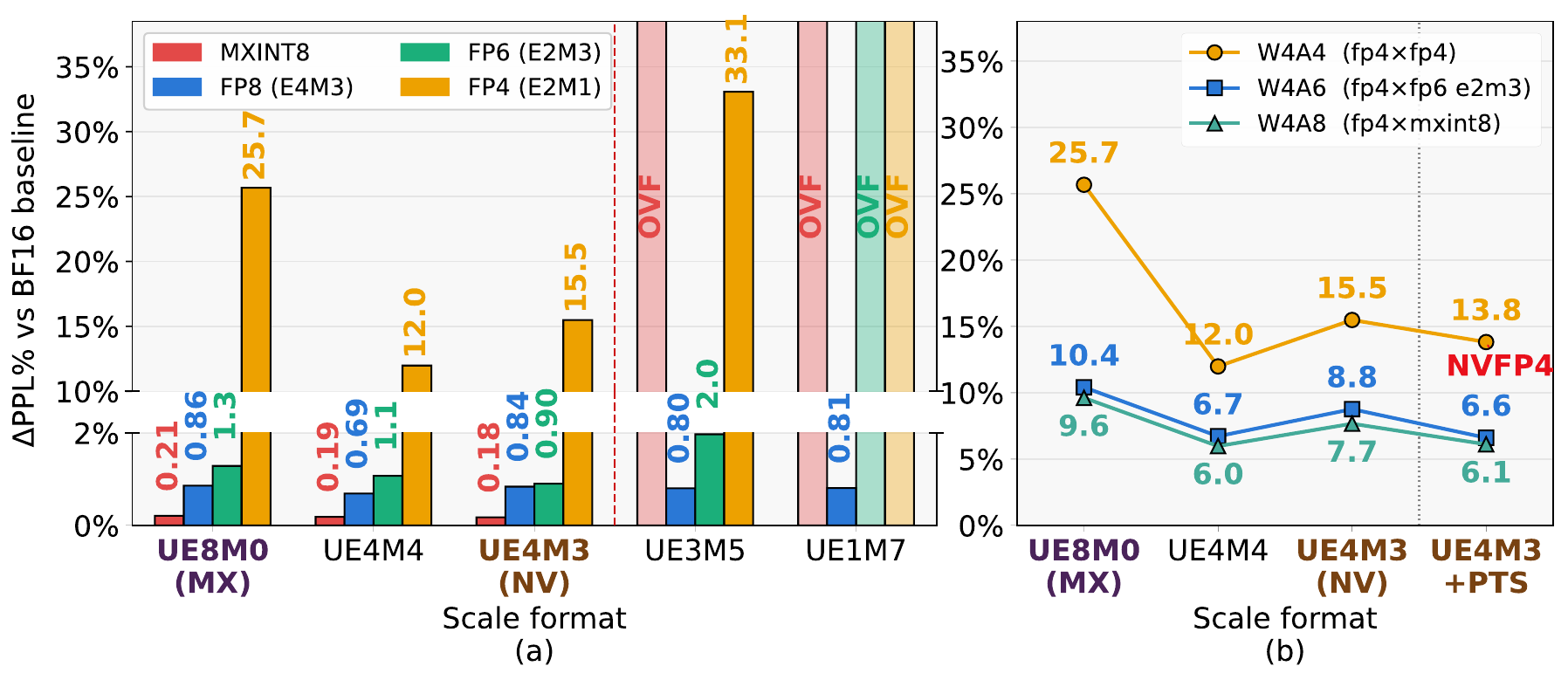}
\Description{Two charts of the relative perplexity loss of Qwen2.5-3B.
(a) Grouped bars for symmetric MXINT8, FP8, FP6, and FP4 formats under the
scale formats UE8M0, UE4M4, UE4M3, UE3M5, and UE1M7. FP4 has the largest loss,
and several formats overflow under UE3M5 and UE1M7. (b) Lines for FP4 weights
with FP4, FP6, and MXINT8 activations under UE8M0, UE4M4, UE4M3, and UE4M3 with
a per-tensor scale. FP6 and MXINT8 activations give a much lower loss than FP4
activations.}
\caption{Relative PPL loss (\%) for (a) symmetric
format configurations (W\,=\,A) and (b) asymmetric
format configurations with FP4 weights across scale formats.
NVFP4 in (b) additionally includes NVIDIA's FP32 PTS.
NV (NVFP4-like) denotes UE4M3 scale without PTS.
OVF (overflow) marks configurations where $\Delta\text{PPL}$ exceeds
40\%.}
\label{fig:scale_dr}
\end{figure}

We define the MX-like design space over the full 8-bit scale format
family, paired with symmetric and asymmetric weight-activation precision.
Unlike NVFP4's UE4M3 scale, which leaves its sign bit permanently unused
for FP8 alignment, this family reallocates every bit to exponent range or
mantissa precision, and we characterize its accuracy landscape to identify
practical configurations.

The element formats evaluated, FP4 E2M1, FP6 E2M3/E3M2, FP8
E4M3/E5M2, and MXINT8, follow the OCP Microscaling (MX)
specification~\cite{ocp_mx_spec_2023}, spanning precisions from 8 bits down to 4 bits, the range most relevant to edge deployment.
The block scale format family is parameterized as UE$e$M$m$, with $e$ exponent bits and $m$ mantissa
bits ($e+m=8$), decoded as:
\begin{equation}
S = \begin{cases}
    2^{e_b -  \mathrm{bias}}\!\left(1 + m_b/2^m\right) & e_b > 0 \\
    2^{1-\mathrm{bias}} \cdot m_b/2^m                 & e_b = 0
\end{cases}, \quad \mathrm{bias} = 2^{e-1}-1.
\label{eq:exmy}
\end{equation}

For UE8M0 ($m=0$), finite scales are decoded as
$S=2^{e_b-127}$ for $e_b\in\{0,\ldots,254\}$.
The encoding $e_b=255$ is reserved for NaN, as defined in \cite{ocp_mx_spec_2023}.
The search
space spans UE8M0 to UE0M8, with UE4M3 additionally included to
enable direct comparison with NVFP4.
For each scale format, the block scale factor is computed as
$S = \max|x| / Q_{\max}$ with round-up, following the block-scaled
quantization procedure adopted in recent MX training and inference
recipes~\cite{mishra2025recipespretrainingllmsmxfp8,tseng2025trainingllmsmxfp4}, where $Q_{\max}$ is the
maximum representable value of the element format.

However, this design space, spanned by scale format, weight format,
and activation format, is combinatorially large, making exhaustive
evaluation with a rigorous experimental setup computationally prohibitive.
To prune this design space to a set of practical configurations and identify clear trends within it, we
characterize its accuracy landscape as shown in Fig.~\ref{fig:scale_dr}, with
Qwen2.5-3B~\cite{qwen2.5} and
Llama-3.2-1B~\cite{grattafiori2024llama3herdmodels}, using
WikiText-2~\cite{merity2016pointer} perplexity on 30 randomly sampled
sequences of 2048 tokens.
All quantization uses a block
size of 16 to ensure fair comparison with the NVFP4 quantization setup.
To measure how quantization degrades accuracy relative to each model's own full-precision baseline, we adopt this relative perplexity metric:
\begin{equation}
    \Delta{\mathrm{PPL}}_{\%} = \frac{\mathrm{PPL} -
    \mathrm{PPL}_{\mathrm{base}}}{\mathrm{PPL}_{\mathrm{base}}}
    \times 100\%,
    \label{eq:dppl}
\end{equation}
where lower values indicate better accuracy retention.

The characterization yields three findings that shape the practical
design space. First, the scale format's exponent width has a hard floor. As Fig.~\ref{fig:scale_dr}(a) shows for Qwen2.5-3B, formats with fewer than 4 exponent bits (UE3M5 to UE0M8) degrade accuracy severely across most weight/activation combinations, because the scale cannot represent the activation tensor's dynamic range. This restricts the practical scale formats from UE8M0 to UE4M4 and UE4M3, ruling out
approximately half of the design space. Second, within this
range, fractional scale formats such as UE5M3 and UE4M4 provide a
measurable accuracy advantage over UE8M0 and NVFP4's UE4M3,
particularly in the FP4 regime.
In contrast, NVFP4 additionally introduces a second-level FP32 per-tensor scale on top of its block scale~\cite{nvidia2026pretraininglargelanguagemodels}. As illustrated in Fig.~\ref{fig:scale_dr}(b), incorporating this PTS yields only marginal accuracy improvement within this range, at the cost of two additional whole-tensor memory traversals for tensor-level scaling. Since choosing a better scale format achieves a larger accuracy gain without this overhead, this work prioritizes scale format selection over per-tensor scaling.
Third, asymmetric weight-activation configurations further improve accuracy over symmetric ones at the same weight precision, an effect that is most pronounced in the FP4 regime, as depicted in Fig.~\ref{fig:scale_dr}(b), motivating support for asymmetric operation in the datapath.

The above findings narrow the practical design space to UE8M0 through UE4M4 without a PTS, and identify fractional scale formats such as UE5M3 and UE4M4 as the most promising candidates for the FP4 regime.
The configurations in the retained space
are further evaluated across four models and two datasets in Sec. \ref{sec:exp} to assess the generalizability
of the observed trends.
However, fractional scales require a multiplier rather than a simple
bit-shift, introducing a hardware cost that we account for in the
datapath design presented next.

\subsection{Datapath Template Implementation}
\label{sec:system_template}
\begin{figure*}[!t]
\centering
\includegraphics[width=0.95\textwidth]{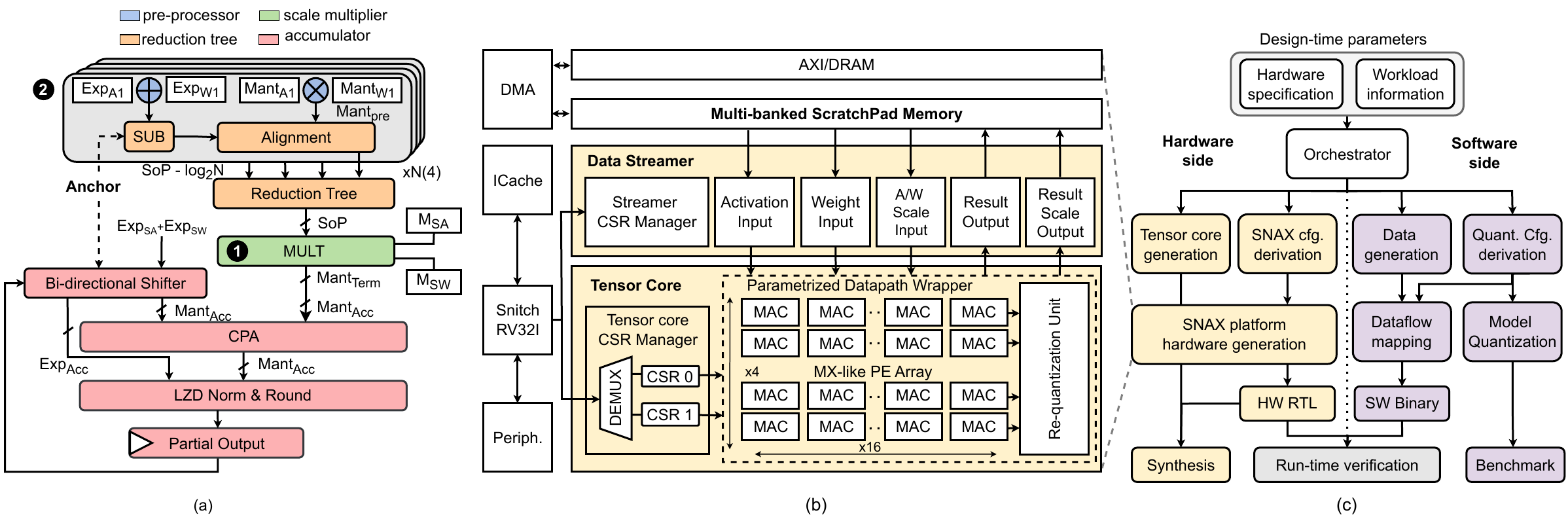}
\Description{(a) Block diagram of the MX-like MAC datapath: exponent addition
and mantissa multiplication of the operand pairs, alignment against an anchor,
a reduction tree, a scale multiplier, a bi-directional shifter, a
carry-propagate adder, and leading-zero normalization with rounding into the
partial output. (b) The SNAX cluster: a Snitch RISC-V core, DMA, a multi-banked
scratchpad memory, data streamers for activations, weights, scales, and
results, and a tensor core with a CSR manager, a 4 by 16 array of MAC units,
and a re-quantization unit. (c) The co-design flow: an orchestrator takes the
design-time parameters and drives the hardware side (tensor core and SNAX
generation, RTL, synthesis) and the software side (data generation, dataflow
mapping, binary, model quantization, benchmark), which meet in the run-time
verification.}
\vspace{-2mm}
\caption{The overall MX-like co-design exploration framework. It consists of (a) an MX-like MAC datapath template within (b) the tensor core integrated into the open-source SNAX platform, and enables agile generation with (c) an automated co-design flow.}
\label{fig:integration}
\vspace{-2mm}
\end{figure*}

To investigate how activation, weight and scale precision affect the hardware implementation, we propose a MAC datapath template.
Fig.~\ref{fig:integration}a shows the datapath. The pre-processor forms four products of operands which a shared reduction tree then sums.
Alignment before the tree is performed against a pre-computed anchor position, which avoids
comparing exponents across products~\cite{lutz2024arithmetic}. A scale multiplier follows, and a bi-directional shifter aligns the partial result with the current scaled term through a carry-propagation adder (CPA). Then the accumulation result goes through a leading-zero detection (LZD) normalizer and round-ties-to-even (RNE) rounder~\cite{ocp_mx_spec_2023}.
The output after the LZD normalizer and rounder is written back to the accumulation register, which is read out again as the running sum for the next MAC cycle.

Prior MX-compatible designs focus on FP8 operands with E8M0 scale~\cite{mxdotp,lutz2024arithmetic}. Instead, to enable our exploration into the MX-like design space, our proposed general hardware template is built on top of it with two innovations:
\cnum{1}  When the scale factor is fractional and cannot be folded into the bi-directional shifter as a power-of-two, we implement an additional multiplier after the reduction tree.
\cnum{2} We parameterize the datapath on five format axes ($\mathrm{e}_\mathrm{A},\mathrm{m}_\mathrm{A},\mathrm{e}_\mathrm{W},\mathrm{m}_\mathrm{W},\mathrm{m}_\mathrm{S}$), representing the exponent and mantissa widths of activations, weights, and shared scales\footnote{The exponent size is $\mathrm{e}_\mathrm{S} = 8-\mathrm{m}_\mathrm{S}$, we also assume the activations and weights share the same scale type. The scale exponent $e_\mathrm{S}$ is absent from the width chain because it enters only through the accumulator bi-directional shifter's control path.} and generate a datapath specialized to each configuration. A Chisel~\cite{chisel_ucberkely}
generator evaluates those parameters at elaboration time and
emits one RTL instance per design point. This parameterization absorbs asymmetric operand pairs and fractional scale, enabling a fair comparison of the hardware cost across the MX-like design space.

Table~\ref{tab:widths} gives each stage's width as a function of these axes. These widths reveal two properties that drive the cost.
First, the operand exponents set the tree
width from both ends. The Anchor sits at the largest product magnitude, and the
Sum-of-Product (SoP) field extends down to the smallest subnormal product ($E^A_{sub}$,
$E^W_{sub}$). Wide-exponent formats therefore pay twice. Second, the scale
multiplier inherits that same SoP width. The cost of a fractional scale is thus
inseparable from the operand pair. Moreover, the $\mathrm{Mant}_{\mathrm{Acc}}$  appends a fixed significand of the partial result (e.g., 8 bits for BF16), which is independent of any format axis.

To make this concrete, consider E4M3\,$\times$\,E2M1 with the UE6M2 scale. The activation E4M3 has maximum normal exponent  $E^A_{\max}=8$ and minimum $\mathrm{E}^\mathrm{A}_{\min}=-6$; the weight E2M1 has $\mathrm{E}^\mathrm{W}_{\max}=2$ and $E^W_{\min}=0$. So the largest possible product magnitude is $2.625 \times 2^{10} \approx 2^{11.4}$~\cite{ocp_mx_spec_2023}, placing its leading bit 11 positions above $2^0$ as the anchor; the smallest subnormal product $2^{-9} \times 2^{-1} = 2^{-10}$ places its leading bit $\bigl|E^A_{sub}+E^W_{sub}\bigr|=10$ positions below. Adding $\log_2 N = 2$ for four-way accumulation and a sign bit lifts the top to $2^{14}$, so the SoP field spans positions $2^{-10}$ through $2^{14}$, 25 bits in total. The scale multiplier then appends $2(m_S+1)+1 = 7$ bits for $\mathrm{Mant}_{\mathrm{Term}} = 32$, and the BF16 accumulator contributes a further 9 bits at the CPA for $\mathrm{Mant}_{\mathrm{Acc}} = 41$.

\begin{table}[t]
\centering
\caption{Datapath widths derived for format configurations. For UE8M0 scale ($m_S$=0) the 4th row $\mathrm{Mant}_{\mathrm{Term}}$ collapses to $\mathrm{SoP}$.}

\label{tab:widths}
\resizebox{0.9\linewidth}{!}{\begin{tabular}{@{}l l l@{}}
\toprule
Signal & Meaning & Formula \\
\midrule
$\mathrm{Mant}_{\mathrm{Pre}}$  & product width of mantissa
  & $(\mathrm{m_A}+1)+(\mathrm{m_W}+1)$ \\[2pt]
$\mathrm{Anchor}$               & anchor-alignment position
  & $(\mathrm{E^A_{max}}+\mathrm{E^W_{max}}+1)+\log_2\mathrm{N}+1$ \\[2pt]
$\mathrm{SoP}$                  & Sum of Product width
  & $\begin{aligned}[t]
      &1+\mathrm{Anchor}+\bigl|E^A_{sub}+E^W_{sub}\bigr|
    \end{aligned}$ \\[2pt]
$\mathrm{Mant}_{\mathrm{Term}}$ & scaled term width
  & $\mathrm{SoP}+2(\mathrm{m_S}+1)+1$ \\[2pt]
$\mathrm{Mant}_{\mathrm{Acc}}$  & accumulator width
  & $\mathrm{Mant}_{\mathrm{Term}}+8+1$ \\
\bottomrule
\end{tabular}}
\vspace{-4mm}
\end{table}

\subsection{Automated Co-design Framework}
\label{sec:dse_framework}
The joint HW-SW DSE requires evaluating hundreds of quantization
configurations, each demanding a matched hardware instance and a
consistent software test setup. Performing this manually
means hand-deriving the run-time values that define memory layout and
dataflow mapping for every configuration, and manually keeping the
software reference and hardware RTL in sync, a process that is both
labor-intensive and error-prone at this scale. We therefore introduce
an automated co-design flow that derives every downstream parameter,
dataflow mapping, and RTL instance from a single design-time input,
without manual calculation.

Our co-design flow builds on the SNAX
platform~\cite{Antonio2025AnOH}, an open-source RISC-V compute cluster in
which a lightweight Snitch~\cite{Zaruba2021Snitch} core offloads
computation to attached accelerators through Control and Status Register
(CSR) instructions. As shown in Fig.~\ref{fig:integration}b, a shared scratchpad
memory and parameterized data streamers absorb the varying data widths
across our DSE, while within the tensor core a CSR manager, a
two-dimensional PE array with design-time configurable spatial
parallelism, and a re-quantization unit carry out the MAC and format
conversion.

Building on this platform, the flow is driven by a top-level
orchestrator, as illustrated in Fig.~\ref{fig:integration}c. The
orchestrator takes a unified set of design-time parameters including operand datatypes, scale datatypes, workload dimension and tensor core dimension, and translates them into the configurations that feed the downstream flow. It then drives two
parallel paths from this shared configuration. On the software compilation path, the orchestrator invokes data generation to
produce quantized test data and software reference outputs by our custom quantization function, along with derivation of run-time CSR values for the target quantization configuration and workload to define the memory layout and computation dataflow. Together with a run-time library, this path yields a software binary. In parallel, the selected LLM is quantized by our custom quantizer and evaluated on the target dataset to obtain the accuracy benchmark.
On the hardware generation path, the tensor core datapath is generated directly according to the orchestrator's parameters, while the derived SNAX
configuration, including the streamer width computed from the target bandwidth requirement, is fed into the SNAX toolchain
to generate the surrounding components such as the streamers and CSR manager. Both steps use Chisel~\cite{chisel_ucberkely} to produce synthesizable HW RTL.
The software binary and hardware RTL are then co-simulated to confirm that the generated hardware behaves consistently with the software reference. Separately, the RTL is synthesized to obtain area and power, which together with the accuracy benchmark form the inputs to the HW-SW co-analysis.

\begin{figure}
    \centering
    \includegraphics[width=\linewidth]{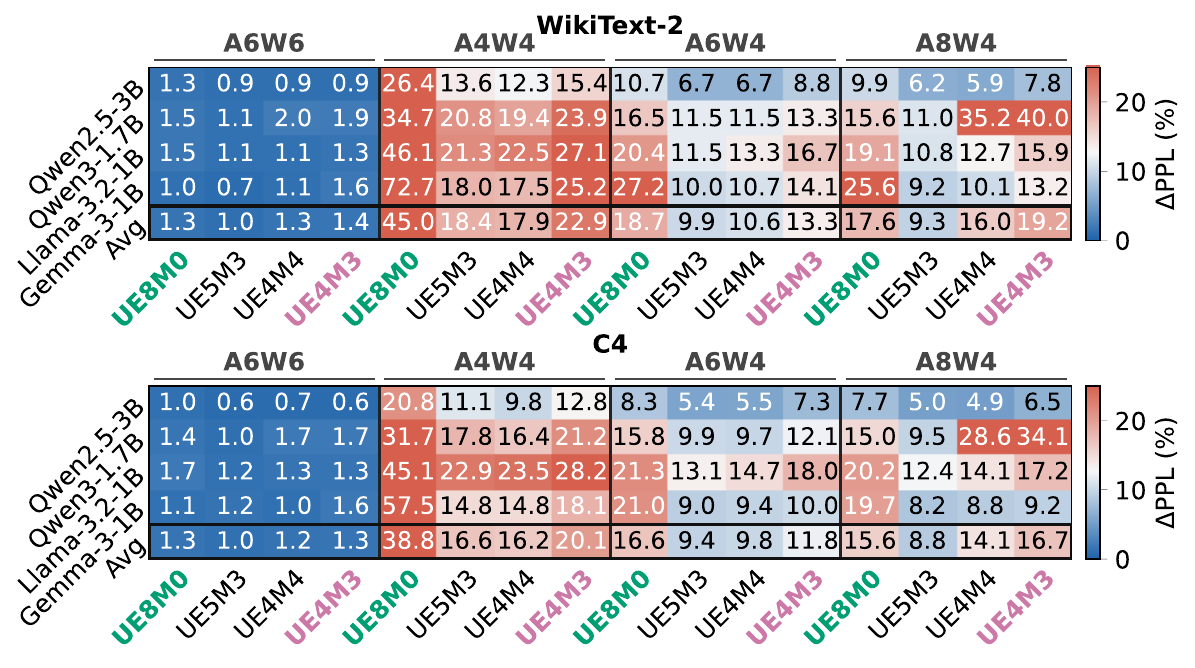}
    \Description{Two heatmaps, for WikiText-2 (top) and C4 (bottom), of the
    relative perplexity degradation of Qwen2.5-3B, Qwen3-1.7B, Llama-3.2-1B,
    and Gemma-3-1B, and of their average. The columns are the scale formats
    UE8M0, UE5M3, UE4M4, and UE4M3 for the A6W6, A4W4, A6W4, and A8W4
    configurations. All values at A6W6 are about 2\% or lower, and the largest
    average is for UE8M0 at A4W4.}
    \vspace{-5mm}
    \caption{Relative PPL degradation across
scale formats and W/A configurations, on WikiText-2 (top) and C4
(bottom). W/A precision: 8 = MXINT8, 6 = E2M3, 4 = E2M1. Green/pink headers mark the MX/NVFP4 baseline configurations.}
    \label{fig:ppl_heatmap}
    \vspace{-5mm}
\end{figure}

\section{Experimental results}
\label{sec:exp}
\subsection{Experiment Setup}
\label{sec:setup}
On the software side, we
extend the open-source MX quantization library of microxcaling~\cite{microsoft2023microxcaling} and
BitMoD~\cite{bitmod} to produce identical numerical results matching the hardware datapath. We evaluate four language
models: Qwen2.5-3B~\cite{qwen2.5}, Qwen3-1.7B~\cite{qwen3technicalreport},
Llama-3.2-1B~\cite{grattafiori2024llama3herdmodels}, and
Gemma-3-1B~\cite{gemma_2025} on the full test split of WikiText-2~\cite{merity2016pointer} and 256 randomly sampled
sequences of 2048 tokens from C4~\cite{raffel2020exploring} to generalize our findings across datasets. Accuracy is measured as
the perplexity degradation $\Delta\text{PPL}_{\%}$ as defined in
Eq.~\eqref{eq:dppl}.

On the hardware side, all generated designs are synthesized at 400\,MHz with Synopsys Design Compiler on a TSMC 28\,nm process, using standard cells at 0.8\,V supply in the typical-typical corner. Power is estimated with Synopsys PrimeTime PX from the switching activity of a gate-level simulation over 2048 MAC operations to stabilize the average activity. At the system level, we instantiate a 4$\times$16 PE array and its re-quantization units in the SNAX cluster~\cite{yi2024opengemm} to capture the full tensor-core cost, driven by a 32$\times$1024 tile of real weight and activation tensors from a representative transformer layer of Qwen2.5-3B~\cite{qwen2.5}, whose distribution we observe to be consistent across layers. The reported area and power include the on-chip
storage and data-movement overheads associated with asymmetric
operand widths and fractional-scale metadata. Every configuration maps to the same 64-PE array at an easily achievable timing constraint of 400 MHz to ensure the minimum area across all designs and therefore sustains an identical throughput of 204.8\,GOPS.

\subsection{Accuracy Characterization}
\label{sec:accuracy_characterization}

\begin{figure}[t]
  \centering
  \includegraphics[width=\linewidth]{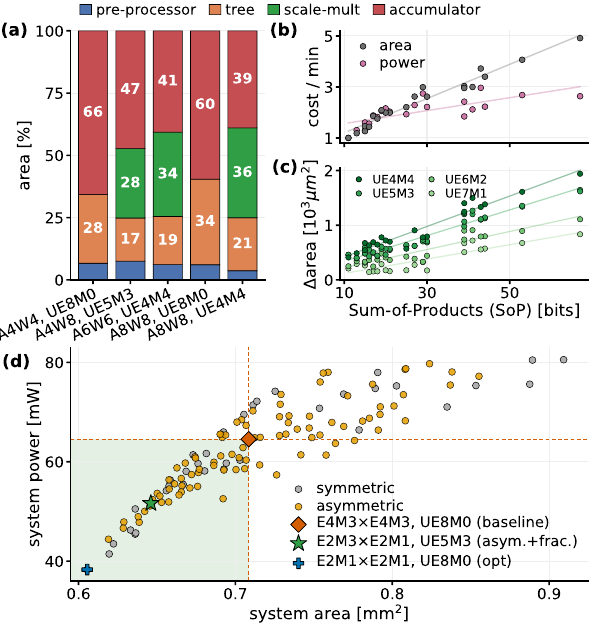}
  \Description{Four panels. (a) Stacked bars of the area share of the
  pre-processor, reduction tree, scale multiplier, and accumulator for five
  format configurations; the accumulator has the largest share. (b) MAC area
  and power, normalized to the MXFP4 configuration, against the
  sum-of-products width at the UE8M0 scale; the area rises faster with the
  width than the power. (c) The extra area of the fractional scales UE4M4,
  UE5M3, UE6M2, and UE7M1 against the sum-of-products width; the extra area
  rises with the width. (d) System power against system area for all
  configurations. The FP8 baseline, a representative asymmetric configuration
  with a fractional scale, and the cost-only optimum are marked, and a shaded
  region holds the configurations below the baseline in both area and power.}
  \vspace{-2em}
  \caption{
\textbf{(a)} Area breakdown (\%) of five formats.
\textbf{(b)} Area and power of all operand pairs at UE8M0, each normalized to the MXFP4 configuration, against the aligned SoP width.
\textbf{(c)} Incremental area by fractional scale against UE8M0.
\textbf{(d)} System-level scatters, where the
shaded region contains settings below the MXFP8 baseline in both area
and power.}
  \vspace{-1em}
  \label{fig:hardware_cost}
\end{figure}

\begin{figure*}[t]
  \centering
    \includegraphics[width=\linewidth]{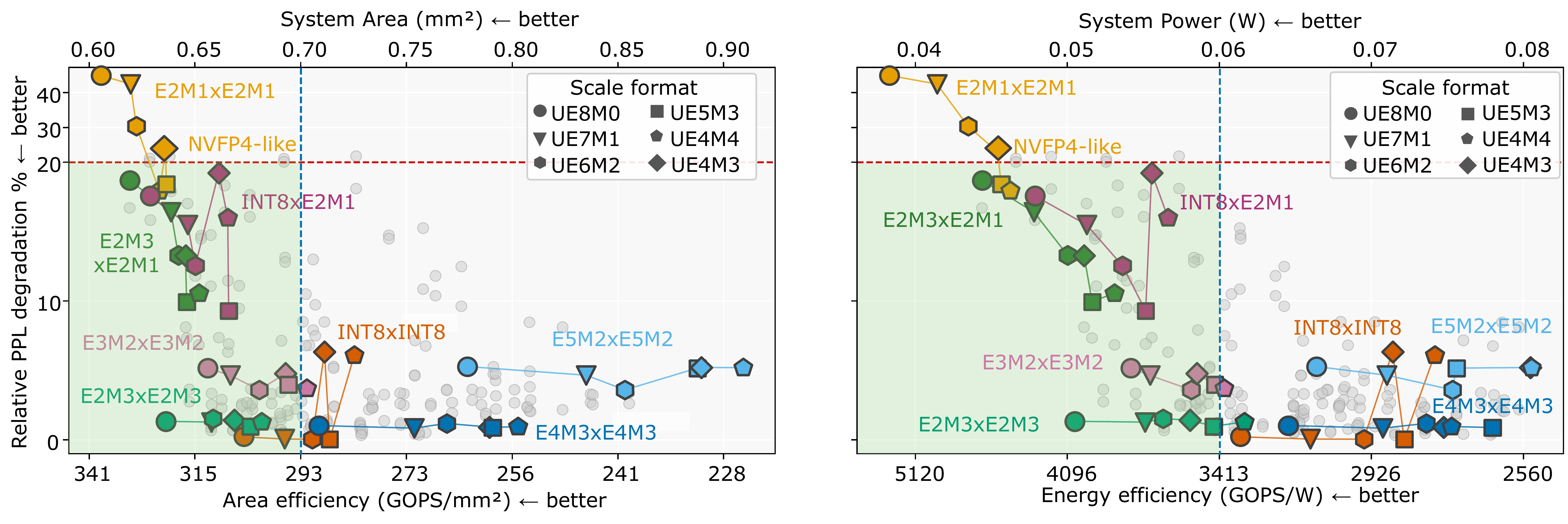}
  \Description{Two scatter plots of the mean relative perplexity degradation
  against the system area (left) and the system power (right) for all
  configurations. The marker shape gives the scale format, and lines connect
  the scale formats of the same operand pair. A horizontal red line marks the
  20\% degradation budget, and vertical blue lines mark the area and power
  budgets. The labeled points include E2M1xE2M1, the NVFP4-like
  configuration, and E2M3xE2M1.}
  \caption{Pareto front of our HW-SW co-design exploration. The x-axis reports the area (left) and power (right) of the NPU system. The y-axis reports mean $\Delta\text{PPL}_{\%}$
    averaged over four models~\cite{qwen2.5,qwen3technicalreport,grattafiori2024llama3herdmodels,gemma_2025} on the full test split of WikiText-2 ~\cite{merity2016pointer}. The red dotted line marks the 20\,\%  mean relative perplexity degradation budget. The blue dotted line marks the specific hardware constraint on edge~\cite{mxdotp,cuyckens2024precision}.}
  \label{fig:pareto}
\end{figure*}

Building on observations in Sec.~\ref{subsec:design_space}, we first evaluate
the UE5M3--UE4M4 divergence and the limits of asymmetric precision across four models.
The effect of scale format on accuracy depends critically on the
target precision regime: it is negligible at high precision but
becomes decisive once quantization is pushed aggressively. Taking
A6W6 as representative of the high-precision regime, all four scale
formats yield $\Delta$PPL$_{\%}$ within a narrow 1.0--1.4\% range
across every model, as Fig.~\ref{fig:ppl_heatmap} shows,
making the choice of scale format largely insensitive at this
precision.

This changes sharply in the FP4 quantization regime. In the extreme A4W4
case, the coarse UE8M0 scale causes severe degradation,
averaging 45.0\%, while UE5M3, UE4M4 and NVFP4's UE4M3 substantially
outperform it, cutting average
$\Delta$PPL$_{\%}$ roughly in half.
UE5M3 and UE4M4 further outperform UE4M3
by approximately 5 percentage points on average, by reclaiming the unused bit in the format, a gap that stems from Eq.~\eqref{eq:exmy}.
Yet the two are not interchangeable,
since UE5M3 widens dynamic range while UE4M4 improves
precision. Fig.~\ref{fig:ppl_heatmap} also shows that this divergence surfaces sharply on Qwen3-1.7B. Its late-layer activation
outliers, once quantized to MXINT8, overflow UE4M4's and UE4M3's
four-bit exponent range, causing a severe degradation, whereas
UE5M3's extended exponent range absorbs the same outliers and stays the
more robust choice.

On top of scale format selection, asymmetric weight-activation
quantization provides a further improvement, particularly
pronounced in the FP4 regime, as anticipated by the distribution
asymmetry noted in Sec.~\ref{subsec:design_space}. Moving from
A4W4 to asymmetric A6W4, which raises only the activation format to
FP6, cuts $\Delta$PPL$_{\%}$ across all four scale formats, nearly
halving UE5M3's degradation from 18.4\% to 9.9\% and UE4M4's from
17.9\% to 8.8\%. Further raising activation precision to MXINT8
(A8W4) trims UE5M3's degradation only slightly further, to 6.1\%,
while UE4M4 and UE4M3 instead rise to 16.0\% and 19.2\% due to their limited exponent range.

\subsection{Hardware Cost Evaluation}
\label{sec:hardwarecost}

In this subsection, we evaluate where an MX-like format spends its silicon cost,
from the datapath stages up to the system level. We first break
the datapath into stages across formats, as
Fig.~\ref{fig:hardware_cost}(a) shows. The accumulator is consistently
the largest, holding 39--66\% of the area. A fractional scale adds
a scale-multiplier that reaches 36\% and that a UE8M0 scale
lacks, a cost traceable to Table~\ref{tab:widths}, where the
multiplier shares the accumulator's $\mathrm{Mant}_\mathrm{Term}$ field.

We then fix the scale at UE8M0 and sweep operand pairs,
tracking area and power against the aligned sum-of-products (SoP)
width growing with $\mathrm{e}$ and $\mathrm{m}$ of
operands, as shown in Fig.~\ref{fig:hardware_cost}(b). Over SoP
from 11 to 67 bits the MAC area grows 4.9$\times$ but power only
2.6$\times$, since area follows the SoP width whereas
power is set by actual switching in the multiplier and scatters
against SoP. The fractional scale's multiplier operates on
this same SoP field, coupling its cost to the operand pair, as presented in Fig.~\ref{fig:hardware_cost}(c). The
heaviest scale, UE4M4, adds about $0.5\times10^{3}\mu m^{2}$ on the
narrow INT8$\times$E2M1 but $1.9\times10^{3}\mu m^{2}$ on the wide
E5M2$\times$E5M2, a 3.6$\times$ spread that no single design point
reveals.

At the system level, where deployment budgets bind, these
mechanisms pay off, as illustrated in Fig.~\ref{fig:hardware_cost}(d).
Of the 126 configurations, 60 draw both less area and power than
the conventional symmetric-FP8 baseline, E4M3$\times$E4M3
with UE8M0, which uses $0.708mm^{2}$ and $64.5mW$, and 30 of these
are asymmetric with a fractional scale. A representative
case, marked as a green star, is E2M3$\times$E2M1 with
UE5M3, using $0.646mm^{2}$ and $51.6mW$, 9\% less area and 20\%
less power, while carrying a fractional scale the default lacks.
However, the cost axis alone cannot decide which affordable format
is worth adopting. The cost-only optimum, marked as a blue
cross, is the FP4 pair at UE8M0, using $0.606mm^{2}$ and
$38.3mW$, and is ruled out by the accuracy characterization in
Sec.~\ref{sec:accuracy_characterization}.

\subsection{Hardware--Software Co-analysis}

We now bring accuracy and hardware cost together to locate
the true Pareto-optimal configurations, using the joint HW-SW
co-analysis in Fig.~\ref{fig:pareto}. Each point is one
configuration of the MX-like space, its vertical position the mean
$\Delta$PPL$_{\%}$ across four models and its horizontal position the system-level
cost, with every instance behind these points generated, verified,
and synthesized by our automated co-design flow.

We find that the two single-sided conclusions each hold in exactly
one precision regime. Above A6W6 the choice is dominated by
hardware cost, as UE8M0 is Pareto-optimal, while a
fractional scale buys no accuracy but adds up to +22\% system power
and +17\% area over the UE8M0 baseline. The FP4 regime inverts
this: the same UE4M4 scale now buys 27 percentage points of
$\Delta$PPL\, from 45.0\% under UE8M0 down to 17.9\%, while
at FP4 operands the same fractional overhead falls to +4.6\% area
and +20.8\% power. Within the fractional-scale family, UE5M3
emerges as the sweet spot, recovering 25.3\% of
$\Delta$PPL relative to the NVFP4-like UE4M3 at only +0.06\% area
and +0.49\% power over that baseline, and showing more robustness
than UE4M4, as discussed in
Sec.~\ref{sec:accuracy_characterization}. UE5M3 is therefore our
recommended FP4 scale format.

We further mark three deployment thresholds in
Fig.~\ref{fig:pareto}: 1) a 20\,\%
perplexity-degradation budget,
guided by
comparisons with smaller unquantized models of the same families
in our experiments, 2) a 60\,mW power budget, corresponding to
3413\,GOPS/W at our constant throughput which is close to the prior precision-scalable MX clusters in FP4
mode~\cite{cuyckens2024precision}, and 3) a 0.7\,mm$^2$ area
budget, in line with prior MX systems at 0.59--0.60\,mm$^2$
on more advanced nodes~\cite{mxdotp,cuyckens2024precision}, with
headroom at the 28\,nm technology we use.
Note that our conclusions are not tied to these specific values, and they hold for any accuracy threshold in 15–25\% and any power budget in 56–64\,mW.
Within them, neither traditional
baseline is feasible. Symmetric MXFP4 fails at 45.0\% and the
per-tensor-scale-free NVFP4-like UE4M3 configuration at 22.9\%,
while most feasible points are asymmetric, fractional-scale
configurations that neither standard has defined before, led by A6W4
(E2M3$\times$E2M1) with UE5M3 scale.

This co-analysis lets a practitioner pick a configuration for
their own accuracy and hardware budget, and points to which
precision combinations future hardware should support as edge
deployment evolves~\cite{wang2025cognitiveedgecomputingcomprehensive,
wang2026activationalignmentthealignmentdiversitytradeoff}.

\section{Conclusion}

In this work, we characterize the previously unexplored joint accuracy-silicon Pareto frontier of the MX-like design space, spanned by scale format, weight format, and activation format. We define the full 8-bit scale format family jointly with asymmetric weight-activation precision, build a per-format specialized MAC datapath template for fair hardware cost evaluation across the space, and integrate both into an automated co-design flow on the open-source SNAX NPU platform. Our joint exploration shows that scale format is largely inconsequential above A6W6, where UE8M0 remains Pareto-optimal, but becomes decisive under FP4, where UE5M3 paired with asymmetric weight-activation precision reduces perplexity degradation by up to 25.3\% relative to the NVFP4-like baseline with only a negligible hardware overhead.

\begin{acks}
The authors would like to thank Ryan Antonio and Xiaoling Yi from ESAT-MICAS, KU Leuven, Belgium for the helpful discussions and supports on this work.
This project has been partly funded by the
\grantsponsor{ERC}{European Research Council (ERC)}{https://doi.org/10.13039/501100000781}
under grant agreement No.~\grantnum{ERC}{101088865}, the Flanders AI Research
Program, and long-term structural Methusalem funding by the
\grantsponsor{FlemishGov}{Flemish Government}{https://doi.org/10.13039/501100011878}.
\end{acks}

\bibliographystyle{ACM-Reference-Format}
\bibliography{ref_simplified_fixed}

\end{document}